\documentstyle{article}
\begin{document}
\newcount\nummer \nummer=0
\def\f#1{\global\advance\nummer by 1 \eqno{(\number\nummer)}
      \global\edef#1{(\number\nummer)}}
\def\nn{\nonumber \\}
\def\be{\begin{equation}}
\def\ee{\end{equation}}
\def\eel#1 {\label{#1}\end{equation}}
\def\ba{\begin{eqnarray}}
\def\ea{\end{eqnarray}}
\def\la{\label}
\def\re{(\ref}
\def\rz#1 {(\ref{#1}) }
\def\i{{\rm i}}
\let\a=\alpha \let\b=\beta \let\g=\gamma \let\d=\delta
\let\e=\varepsilon \let\ep=\epsilon \let\z=\zeta \let\h=\eta
\let\th=\theta
\let\dh=\vartheta \let\k=\kappa \let\l=\lambda \let\m=\mu
\let\n=\nu \let\x=\xi \let\p=\pi \let\r=\rho \let\s=\sigma
\let\t=\tau \let\o=\omega \let\c=\chi \let\ps=\psi
\let\ph=\varphi \let\Ph=\phi \let\PH=\Phi \let\Ps=\Psi
\let\O=\Omega \let\S=\Sigma \let\P=\Pi \let\Th=\Theta
\let\L=\Lambda \let\G=\Gamma \let\D=\Delta

\def\w{\wedge}
\def\0{\over } \def\1{\vec } \def\2{{1\over2}} \def\4{{1\over4}}
\def\5{\bar } \def\6{\partial }
\def\7#1{{#1}\llap{/}}
\def\8#1{{\textstyle{#1}}} \def\9#1{{\bf {#1}}}

\def\({\left(} \def\){\right)} \def\<{\langle } \def\>{\rangle }
\def\[{\left[} \def\]{\right]} \def\lb{\left\{} \def\rb{\right\}}
\let\lra=\leftrightarrow \let\LRA=\Leftrightarrow
\let\Ra=\Rightarrow \let\ra=\rightarrow
\def\ul{\underline}

\let\ap=\approx \let\eq=\equiv 
        \let\ti=\tilde \let\bl=\biggl \let\br=\biggr
\let\bi=\choose \let\at=\atop \let\mat=\pmatrix
\def\CL{{\cal L}} \def\CD{{\cal D}} \def\rd{{\rm d}} \def\rD{{\rm D}}
\def\CH{{\cal H}} \def\CT{{\cal T}} \def\CM{{\cal M}} \def\CI{{\cal I}}
\def\CX{{\cal X}} \def\CA{{\cal A}} \def\CS{{\cal S}}
\newcommand{\dR}{\mbox{{\sl I \hspace{-0.8em} R}}}
\begin{titlepage}
\renewcommand{\thefootnote}{\fnsymbol{footnote}}
\renewcommand{\baselinestretch}{1.3}
\hfill  TUW - 94 - 03 \\
\medskip
\hfill hep-th/9405110\\
\medskip
\vfill

\begin{center}
{\LARGE {Poisson Structure Induced (Topological) Field Theories}
 \\ \medskip  {}}
\medskip
\vfill

\renewcommand{\baselinestretch}{1} {\large {PETER
SCHALLER\footnote{e-mail: schaller@email.tuwien.ac.at} \\
\medskip THOMAS STROBL\footnote{e-mail:
tstrobl@email.tuwien.ac.at} \\ \medskip\medskip
\medskip \medskip
Institut f\"ur Theoretische Physik \\
Technische Universit\"at Wien\\
Wiedner Hauptstr. 8-10, A-1040 Vienna\\
Austria\\} }
\end{center}

\vfill
\renewcommand{\baselinestretch}{1}                          

\begin{abstract}
A class of two dimensional field theories, based on (generically
degenerate) Poisson structures and generalizing gravity-Yang-Mills
systems, is presented. Locally, the solutions of the classical
equations of motion are given. A general scheme for the
quantization of the models in a Hamiltonian formulation is found.
\end{abstract}

\vfill
\hfill Vienna, May 1994  \\
\end{titlepage}

As we will show in the present letter, any Poisson structure on
any finite dimensional manifold naturally induces a two
dimensional topological field theory. Further nontopological
theories are obtained by adding nontrivial Hamiltonians to the
topological theories. Pure gravity and gauge theories in two
dimensions may be seen as special cases of this much larger class
of models.

Choosing special coordinates on the Poisson manifold $N$,
the equations of motions of the models associated to the Poisson
structure on $N$ can be solved locally.

The Poisson structure generates a foliation of
$N$ into symplectic leaves and induces a connection in the loop
space over each of these leaves. Reinterpreting the constraints
of the associated model as horizontality conditions on complex
line bundles over these loop spaces, a general scheme for the
quantization of the models in a Hamiltonian formulation
(restricting the topology of the space time manifold to the one
of a cylinder) is obtained.
It yields finite dimensional quantum mechanical systems, generically
including discrete degrees of freedom of topological origin.

Denote by $M_3$ a three dimensional manifold, whose boundary
$M=\partial M_3$ is the world sheet of a two dimensional field
theory. Denote by $N$ a manifold of arbitrary dimension with a
(generically degenerate) Poisson structure. The latter
may be
expressed in terms of an antisymmetric tensor field $P\in \wedge^2TN$.
Likewise $P$ may be regarded as a map
$T^ *N \to TN$ mapping a one-form to its contraction with
the tensor field. As a consequence of the Jacobi identity
the flow of the
Hamiltonian vector field $V_f=P($d$f,\cdot )$ associated to
an arbitrary function $f$ on $N$ leaves the Poisson
structure invariant. Furthermore, the Hamiltonian
vector fields
form a closed Lie
algebra corresponding to an
(infinite dimensional) subgroup $G_H$ of the P-morphism group.

Let
\be
 \begin{array}{rl}
  (N,P)\hookrightarrow &K\\ &\;\downarrow\\ &M_3
 \end{array}
\eel fibbu
be a fiber bundle over $M_3$ with fiber $N$ and $G_H$ as the
structure group. Let $A$ be a connection on $K$; denote by $D$
and $F$ the covariant derivative and the curvature,
respectively.
As locally the connection is one-form valued on $M$ and a
Hamiltonian vector field on $N$, we have in
coordinates $x^\mu$ on $M_3$ and $X^i$ on $N$
\be
 P={1\over 2} P^{ij}\6_i\wedge\6_j \, , \quad
 A=A_{\mu ,i}dx^\mu P^{ij}\6_j
\eel coord
where we used the abbreviations
\be
 \6_i={\6 \over \6X^i} \, ,\quad A_{\mu,i}=\6_iA_\mu \, .
\ee
Expressing the square of the covariant derivative via a Hamiltonian
vectorfield, the curvature is calculated:
\be
 \begin{array}{c}
   D=d+A \, , \quad D\wedge D=F_{,i}P^{ij}\6_j \, ,\\[2pt]
   F = (A_{\nu,\mu}+{1\over 2}A_{\mu,i}A_{\nu,j}P^{ij}+C_{\mu\nu} )
   dx^\mu \wedge dx^\nu \, ,\quad C_{\mu\nu,i}P^{ij} = 0 \, .
 \end{array}
\eel curva
The appearance of the $C_{\mu\nu}$ reflects an ambiguity in
the definition of $F$.

With curvature and covariant derivative it is straightforward
to formulate a $G_H$ invariant action:
\be
 L=\int_{M_3}F,{_i}\wedge DX^i \, .
\eel acti3
If $d\,C_{\mu\nu} dx^\mu \wedge dx^\nu =0 $, the integrand is exact and
the action may be rewritten as a functional of fields
$A_i(x)$, $X^i(x)$ on $M$ according to
\be
 \begin{array}{c}
   L=\int_{M=\6M_3}A_i\wedge dX^i+{1\over 2} P^{ij}A_i\wedge A_j + C
      \, , \\[2pt]
   A_i(x)=A_{\mu,i}(x,X(x))dx^\mu \, ,
      \quad C(x,X)=C_{\mu\nu} dx^\mu \wedge dx^\nu
 \end{array}
\eel acti2
It is invariant under the symmetries induced by
$G_H$-transformations
\be
 \d_\ep X^i =  \ep_i(x) P^{ij} \, , \quad
\d_\ep A_i = d\ep_i + {P^{lm}}_{,i} A_l \ep_m
\eel symme
up to a total divergence only.

Generically, there is no natural choice of $C$
except for $C=0$, yielding a topological field
theory.\footnote[1]{In this case an
action equivalent to \rz acti2
has been proposed independently in \cite{Ike}.}
Given a volume-form $\e$
on $M$ and a $G_H$-invariant function $\tilde C$ on $N$, however,
a possible choice of $C$ is given by $C=\e\tilde C$.


{}From \rz acti2 the equations of motion follow immediately:
\be
 \begin{array}{c}
  dX^i +  P^{ij} A_j=0 \, , \\[2pt]
  dA_i + \2 P^{lm}{}_{,i} A_l \wedge A_m + C_{,i}=0 \, .
 \end{array}
\eel equom

By construction the action is invariant under diffeomorphisms of $M$
for $C=0$. Indeed, in this case a
diffeomorphism in the direction of a vectorfield $\xi\in TM$ is
generated on-shell by the field dependent
choice $\ep=i_\xi A$ in (\ref{symme}).

Though $P$ is degenerate in general, it defines a bijective map
$T^*N/$Ker$(P)\to \,$Im$(P)$.
The Jacobi identity guarantees Im$(P)$ to be an involutive
system of vector fields and (by the Frobenius theorem) to generate an
integral surface S
(symplectic leave) through any point $X\in N$
with $TS \sim $Im$(P)$.
For the antisymmetry of the Poisson
structure Ker$(P)$ is orthogonal to Im$(P)$. On $S$ this allows the
identification $T^*S \sim T^*N/$Ker$(P)$.
So $P$ induces a bijective map $T^*S \to TS$. By virtue of the
Jacobi identitity the two-form
$\Omega_S \in T^*S\wedge T^*S$ associated with the inverse of this map
is closed and thus symplectic.

Excluding singular points of $P$, we may always find local coordinates
(Casimir-Darboux coordinates)
$(X^I,X^\alpha)$, $I=1,..,$dim$($ker$(P))$, such that the symplectic
leaves are described by $X^I=const$ and, in terms of the $X^\alpha$,
$\Omega_S$ is in Darboux form for each $S$.
In these coordinates $C$ is a function of the $X^I$ only and
the equations of motion simplify to
\be
 dX^I=0 \, , \quad
 dA_I = -C_{,I} \, , \quad
 A_\alpha = (\O_S)_{\beta\alpha} dX^\beta \, .
\eel eomcd
So the
$X^I(x)$ have to be constant on $M$, but otherwise arbitrary, whereas
the $X^\alpha(x)$ remain completely undetermined by the field
equations.
Any choice of the latter determines $A_\alpha$ uniquely through
the last equation in (\ref{eomcd}).
Each of the $A_I$ is determined up
to an exact one-form only.

Still one has not made use of the gauge freedom. As is obvious from
\rz symme any choice of the $X^\alpha$ is gauge equivalent,
and also $A_I \sim A_I
+ dh_I$, where the $h_I$ are arbitrary functions.
Thus locally any solution to the field equations is uniquely
determined by the constant values of the $X^I$.

Additional structures evolve, if global apects are
taken into account. These will be dealt with elsewhere.

For a Hamiltonian formulation of the theory let us assume $M$ to be of
the form $S^1\times R$ parametrized by a 2$\pi$-periodic coordinate
$x^1$ and the evolution parameter $x^0$. One then finds the zero
components $A_{0i}$ of the $A_i$ to play the role of Lagrange
multipliers giving rise to the system of first class constraints
($\6=\6 /\6 x^1$)
\be
 G^i\equiv \6X^i +  P^{ij} A_{1j} \approx 0 \, .
\eel const
The fundamental Poisson brackets are given by
$ \{ X^i(x^1), A_{1j}(y^1) \} = - \d^i_j \d(x^1-y^1)$
and the Hamiltonian reads ($C =\hat C dx^0dx^1$)
\be
  H= \int dx^1 (\hat C  - A_{0i}G^i) \, .
\eel hamil

To quantize the system in an $X$-representation we consider quantum
wave functions as complex valued functionals on the space $\Gamma_N$
of parametrized smooth loops in $N$:
\be
 \Gamma_N=\{ \CX : S^1\to N,x\to X(x) \} \, .
\eel loops
Following the Dirac procedure, only such quantum states are
admissible which satisfy the quantum constraints
\be
 \hat G^i(x) \Psi[\CX ] = \left(
 \6 X^i(x) + i\hbar P^{ij}(X) {\d \0 \d X^j(x)} \right)
 \Psi[\CX ] =0 \, .
\eel quanc

To find the solution of (\ref{quanc}, let us again start with
Casimir Darboux
coordinates: The $I$-components of the constraints $\partial
X^I\,\Psi(\CX )$ restricts the support of $\Psi$ to loops which are
contained entirely in some symplectic leave $S$.

Let us denote the
space of symplectic leaves by $\cal S$.
For $S\in \cal S$ let $\Gamma_S$ be the space of loops on
$S$.
The ansatz $\Psi\vert_{\Gamma_S}=\exp\Phi$ for the restriction of
$\Psi$
to $\Gamma_S$ allows to rewrite the
remaining constraint equations according to
\be
 {\cal A}=-i\hbar \d\Phi \, ,
\eel horco
where $\d$ denotes the exterior derivative on $\Gamma_S$
and $\cal A$ is the one-form on $\Gamma_S$ given by
\be
 \CA = \int_{S^1} \d X^\a(x)({\O_S})_{\a\b}(X(x)) \6 X^\b(x)dx \, .
\eel u1con
(Reinterpreting $\cal A$ as a connection in a $U(1)$-bundle over
$\Gamma_S$ the constraint may be seen as a horizontality condition on
the section  $\Psi\vert_{\Gamma_S}$.)

Equation \rz horco seems to suggest that $\cal A$ has to be exact.
This is not
true, however, as $\Phi$ is determined up to an integer multiple of
2$\pi$ only. Therefore, $\Psi$ is well defined, iff ${\cal A}$ is
closed
and integral, i.e. the intgral of ${\cal A}$ over any closed loop in
$\Gamma_S$ is an integer multiple of $2\pi \hbar$.
To reformulate \rz u1con in a coordinate independent way consider a
path $\gamma$ in $\Gamma_S$. As $\gamma$ corresponds to a one
parameter family of loops in $S$, it spans a two dimensional surface
$\sigma (\gamma )$. $\cal A$ may now be defined alternatively via
\be
 \int_\gamma \CA = \int_{\s(\g)}\Omega_S \, .
\eel altda
As any closed loop in $\Gamma_S$ generates a closed surface in $S$,
$\cal A$ is closed and integral, iff $\Omega_S$ is closed and integral.
The first condition holds, as $\Omega_S$ is symplectic. The integrality
condition, however, may yield a restriction of the support of $\Psi$
to loops over elements of a (possibly discrete) subset of $\cal S$,
if the second homotopy of
the symplectic leaves is nontrivial. For $S$ in this subset, $\Psi$
is determined up to a multiplicative constant on any connected
component of $\Gamma_S$.
As the space of connected components of $\Gamma_S$ is in one to one
correspondence with the first homotopy group of $S$, we may identify
physical states with complex valued functions on $\cal I$ defined via
\be
 {\cal I}=\bigcup_{S\in \tilde \CS} \Pi_1(S) \quad
 \tilde \CS =\{ S\in \CS : \O_S\,
  \hbox{integral}\} \, .
\eel sphst
The Hamiltonian (\ref{hamil}), being constant on each of the symplectic
leaves, defines a function on $\cal I$ and thus becomes a
multiplicative operator upon quantization.
Obviously these results are coordinate independent, although we
intermediately used Darboux coordinates to derive them.

It would be interesting to reproduce the  quantum mechanical system
obtained above via a BRST
cohomology. This is out of the scope of the present letter.
Here we only want to mention that
(despite the appearance of structure functions ${P^{ij}{}_{,k}}$ in the
constraint algebra) the BRS charge takes the minimal
form
\be
  Q= c_i G^i - \2 c_i c_j P^{ji}{}_{,k} b^k \, ,
\ee
where $(c_i,b^i)$ are canonically conjugate fermionic ghost
variables. The nilpotency of $Q$ follows from  the
Jacobi identity.

Let us illustrate our formalism by the example of nonabelian
gauge theories (cf.\ also \cite{Ama}).  There $N$ is the dual
space $g^*$ of the Lie algebra $g$ of the gauge group.
The Lie product on $g$ naturally induces a Poisson structure on
$g^*$. The symplectic leaves generated by the latter are the
coadjoint orbits, i.e. the orbits generated by the
action of the gauge group in $g^*$ equipped with the
standard symplectic form (Kirillov form) \cite{KoS}. The
coadjoint orbits are quantizable spaces, iff this symplectic
form is integral and their quantization yields the
unitary irreducible representations of $g$. This observation
establishes a connection between our representation and the
connection representation of quantum mechanics for non
abelian gauge theories on a cylinder: In the connection
representation, where wave functions are functionals on the
space of gauge connections, the physical wave functions
(i.e.\ the kernel of the constraints) can be identified with
functions on the space of unitary irreducible
representations of $g$ \cite{Raj}.

Choosing a three dimensional linear space $N$, gravitational theories
are
constructed from \rz acti3 via the identification
\be
 A_a = e_a \, ,\quad A_3 = \omega \, ,\quad a=1,2
\eel ident
where $e_a$ and $\omega$ are the zweibein and the spin connection of
the gravity theory. The coordinates $X^a$, $X^3$ become vector valued
and scalar valued functions, respectively, living on the two
dimensional manifold $M$ with the metric $g=e^a e^b \kappa_{ab}$.
The latter is of Euclidian or Minkowski type, corresponding to the
respective signature of the frame metric $\kappa_{ab} = diag
(1,\pm 1)$.

The action of a gravity theory has to be invariant under
diffeomorphisms on $M$ and under frame rotations. The first condition
yields $C=0$ in (\ref{acti2}).
The general solution to the Jacobi identity in three dimensions is
given by \cite{ESI}
\be
 P^{ij} = f\e^{ijk}Q_{,k} \, ,
\eel genps
where $\e$ is the antsymmetric symbol and $f$ and $Q$ are
arbitrary functions on $N$.
Invariance under frame rotations restricts $f$ and $Q$ to be functions
of $X^3$ and $(X)^2=\kappa_{ab}X^aX^b$.
Thus the gravity action we propose is given by \rz acti2 with the
Poisson structure defined by (\ref{genps}), where $f$ and $Q$ are
subject to the above restriction.

A class of examples is provided by
\be
 P^{ij} = \e^{ijk}u_k \, ,
\eel speps
where $u_a = \kappa_{ab}X^b$
and $u_3 $ is an arbitrary function of $X^3$ and $(X)^2$. (It is
somewhat cumbersome, to derive \rz speps from (\ref{genps}). It is
straightforward, however, to convince oneself that \rz speps obeys
the Jacobi identity).
With \rz speps the action \rz acti2 takes the form
\be
 L_{Gr} = \int [X^a(de_a+{\e_a}^b\omega\wedge e_b)
        +X^3d\omega +u_3(e_1\wedge e_2 )] \, .
\eel actgr
Obviously the torsion $T_a=de_a+{\e_a}^b\omega\wedge e_b$
vanishes on shell,
if $u_3$ depends on $X^3$ only.
For some special choices of $u_3$ one may integrate out the $X$-fields
to reformulate $L_{Gr}$ in terms of purely geometric quantities. E.g.,
$u_3=(X^3)^2-\lambda$ leads to the action of torsionless
$R^2$ gravity.

A more detailed analysis of these and further examples
is found in \cite{lnp}, where many of the ideas underlying this
paper have already been presented in a less abstract form.

\section*{Acknowledgement}
We are grateful to A.\ Alekseev for discussions and
valuable comments and to S.\ Shabanov for collaboration on related
topics.

\end{document}